\documentclass[12pt]{article}
\usepackage[dvips]{graphicx}

\begin{document}

\begin{flushright}
KANAZAWA-06-18\\
December, 2006
\end{flushright}

\begin{center}
{\Large \bf Anomalies of Discrete Symmetries and\vspace{6pt}\\
Gauge Coupling Unification}
\end{center}

\begin{center}
Takeshi Araki 
\vspace{6pt}\\
{\it Institute for Theoretical Physics, Kanazawa University, 
Kanazawa 920-1192, Japan}
\end{center}

\begin{abstract}
The anomaly of a discrete symmetry is defined as the Jacobian of
the path-integral measure.
Assuming that the anomaly at low energies is cancelled by the
Green-Schwarz (GS) mechanism at a fundamental scale, we investigate
possible Kac-Moody levels for anomalous discrete family symmetries.
As the first example, we consider discrete abelian baryon number and
lepton number symmetries in the minimal supersymmetric standard model 
with the see-saw mechanism, and we find that the ordinary unification 
of gauge couplings is inconsistent with the GS conditions, indicating 
the possible existence of further Higgs doublets.
We consider various recently proposed supersymmetric models with a 
non-abelian discrete family symmetry.
In a supersymmetric example with $Q_{6}$ family symmetry, the GS 
conditions are such that the gauge coupling unification appears
close to the Planck scale.
\end{abstract}

\newpage

\section{Introduction}
Though the standard model (SM) is very successful, it possesses 
many unsatisfactory features.
One of them is the redundancy of the free parameters in the 
Yukawa sector.
There exist (infinitely) many physically equivalent Yukawa matrices 
that can produce the same physical quantities, such as the fermion 
masses and mixings.
Since there is no principle to fix the Yukawa structure in the SM,
the SM must be extended to impose a family symmetry in order to reduce 
the redundancy of the parameters.
Recently, non-abelian discrete symmetries have been used 
to extend the SM.\cite{A4,S3,D4,Q6,D7,D5,AS4}

Another unsatisfactory feature of the SM is the fine-tuning problem
of the Higgs mass, which can be softened by supersymmetry (SUSY).
Moreover, successful gauge coupling unification in the minimal 
supersymmetric standard model (MSSM) seems to support the existence 
of low-energy SUSY.
However, the situation is not this simple, because SUSY is broken at 
low energies.
It is widely believed that to maintain the nice renormalization 
property of a supersymmetric theory, SUSY must be broken softly.
It is known that there are more than one hundred soft breaking terms
and, unless they are fine-tuned, they create large flavor changing 
neutral currents (FCNC) and {\it CP} violations.
Fortunately, however, this SUSY flavor problem can be softened by 
a non-abelian family symmetry.\cite{susy-f}

The above cited facts suggest that a family symmetry could cure 
certain pathologies of the SM and MSSM.
This leads us to conjecture that a family symmetry consistent
at low energies is a remnant of a symmetry of a more fundamental 
theory.
If this is the case, the symmetry should be anomaly-free, at least
at a fundamental scale.
This provides the motivation to investigate anomalies of the
discrete family symmetries of recently proposed models.
In this paper, we assume that the anomaly of a discrete symmetry at 
low energies is canceled by the Green-Schwarz (GS) mechanism
\cite{GS,4dim-GS,U(1)a} at a more fundamental scale.
In this scenario, the Kac-Moody $k_i$ levels play an important role, 
and if the Kac-Moody levels assume non-trivial values, the 
GS cancellation conditions of anomalies modify the ordinary
unification of gauge couplings.
Note that it is impossible to construct a realistic, renormalizable 
model with a low-energy non-abelian discrete family symmetry
in the case of the minimal content of the $SU(2)_{L}$ Higgs fields.
However, an extension of the Higgs sector may spoil the successful
gauge coupling unification of the MSSM.
Therefore, a possible change of the ordinary unification condition
because of the existence of nontrivial Kac-Moody $k_i$ levels is 
consistent with the assumption regarding the existence of a 
low-energy non-abelian discrete family symmetry.

The anomalies of discrete symmetries have been studied in
Refs.~\cite{IR,kurosawa,babu,dreiner}.
In those papers, it is assumed that all discrete symmetries at low
energies are gauged at high energies, as in the case of Ref.~\cite{DGS}.
In other words, it is assumed that to survive quantum gravity effects, 
such as wormholes,\cite{q-gravity} \ all low-energy discrete symmetries 
must be generated from a spontaneous breakdown of continuous gauge 
symmetries.

In \S2, using Fujikawa's method, \cite{fujikawa} \ we calculate the 
Jacobian of the path-integral measure of an anomalous abelian discrete 
symmetry.
We do this to recall and to demonstrate that the Jacobian can be 
calculated for a finite discrete transformation parameter.
In \S3, we use the result of \S2 to calculate the Jacobian 
for a non-abelian discrete transformation.
Anomaly cancellation is studied in \S5.
First we recall the case of an anomalous $U(1)$, and then we extend the 
cancellation mechanism to the case of discrete symmetries.
In contrast to the treatment of Ref.~\cite{babu}, we do not assume that 
the discrete symmetry in question is not a remnant of a spontaneously 
broken continuous symmetry.
As the first example, we consider discrete abelian baryon number and 
lepton number symmetries in the minimal supersymmetric standard model
with the see-saw mechanism in \S6.
We find that the GS cancellation conditions can be satisfied if 
$k_3/k_2=\mbox{even}/\mbox{odd}$.
This implies that the ordinary unification of gauge couplings is 
inconsistent with the GS conditions, indicating the possible existence 
of further Higgs doublets.
We investigate the unification of gauge couplings for $k_3/k_2=2$ and 
find that the unification scale appears close to the Planck scale for 
three pairs of $SU(2)_{L}$ doublet Higgs supermultiplets.
Recently developed models with $D_7, A_4$ and $Q_6$ symmetries are 
treated in \S7, and \S8 is devoted to a summary.

\section{Anomalies of Abelian discrete symmetries}
An anomaly is a violation of a symmetry at the quantum level.
In the case of a continuous symmetry, an anomaly implies that the 
corresponding Noether current is not conserved.
For discrete symmetries, however, we cannot define an anomaly in this 
way, because there are no corresponding Noether currents.
However, Fujikawa's method,\cite{fujikawa} \ which is based on the 
calculation of the Jacobian of the path-integral measure, can be used 
to define the anomalies of discrete symmetries.
As we see below, the calculation used in this method is basically the 
same as that of the conventional method.

Let us start by considering a Yang-Mills theory with massless fermions
$\psi$ in Euclidean space-time, which can be described by the
following path-integral with the Lagrangian:
\begin{eqnarray}
 &&Z=\int{\cal D}{\bar \psi}{\cal D}\psi
    \ \exp\left[\int d^{4}x{\cal L}\right],\\
 &&{\cal L}=i{\bar \psi}\ /\hspace{-0.3cm}{D}\psi-\frac{1}{2g^{2}}
            {\rm Tr}[F^{\mu\nu}F_{\mu\nu}], \\
 &&D_{\mu}=\partial_{\mu}-iA_{\mu},\\
 &&F_{\mu\nu}=\partial_{\mu}A_{\nu}-\partial_{\nu}A_{\mu}
               -i[A_{\mu}\ A_{\nu}],\\
 &&A_{\mu}\equiv gT^{a}A^{a}_{\mu}\ ,
   \ \ \ {\rm Tr}[T^{a}T^{b}]=\frac{1}{2}\delta^{ab},
\end{eqnarray}
here we have dropped the path-integral measure of the gauge boson 
$A_\mu$, because it does not contribute to the anomaly.
Then we carry out the chiral phase rotation
\begin{eqnarray}
 \psi\rightarrow\psi^{'}=e^{i\alpha\gamma_{5}}\psi,\label{eq,1}
\end{eqnarray}
where $\alpha$ is a finite discrete parameter.
Under this transformation, the Lagrangian is invariant.
Next we consider the transformation properties of the path-integral 
measure.
To this end, we follow Ref.~\cite{fujikawa} and define the eigenstates 
$\varphi_{n}(x)$ of $ /\hspace{-0.3cm}{D}$, i.e.,
\begin{eqnarray}
   /\hspace{-0.3cm}{D}\varphi_{n}(x)=\lambda_{n}\varphi_{n}(x),
   \hspace{1cm}
   \varphi^{\dagger}_{n}\ /\hspace{-0.3cm}{D}
                 =\lambda_{n}\varphi^{\dagger}_{n}(x),
\end{eqnarray}
through the relations
\begin{eqnarray}
 &&\psi_{i}(x)=\sum_{n}a_{n}\varphi_{n,i}(x),\hspace{0.8cm}
 \bar{\psi_{i}}(x)=\sum_{n}\bar{b}_{n}\varphi^{\dagger}_{n,i}(x),
 \hspace{0.6cm}\\
 &&\int d^{4}x\ \varphi_{n,i}^{\dag}(x)\varphi_{m,i}(x)=\delta_{nm},\\
 &&\sum_{n}^{\infty}\varphi_{n,i}(x)\varphi_{n,j}^{\dag}(y)
         =\delta_{ij}\delta^{4}(x-y),\label{eq,28}
\end{eqnarray}
where $i$ and $j$ are spinor indices.
The Jacobian of the path-integral measure for the above transformation, 
which is defined as
\begin{eqnarray}
 {\cal D}\bar{\psi}{\cal D}\psi
           \rightarrow{\cal D}\bar{\psi}^{'}{\cal D}\psi^{'}
           =\frac{1}{J}{\cal D}\bar{\psi}{\cal D}\psi,
\end{eqnarray}
can be written as
\begin{eqnarray}
  J^{-1}=\left\{\det\int d^{4}x\ \varphi_{n,i}^{\dag}(x)
              \left[e^{i\alpha\gamma_{5}}\right]_{ij}
                         \varphi_{m,j}(x)\right\}^{-2}
        \equiv[\det\ C_{nm}]^{-2},
\end{eqnarray}
where
\begin{eqnarray}
C_{nm} &=& \int d^{4}x\ \varphi_{n,i}^{\dag}(x)
         \left[e^{i\alpha\gamma_{5}}\right]_{ij}
         \varphi_{m,j}(x).
 \end{eqnarray}
The quantity $C_{nm}$ is defined as the expansion
\begin{eqnarray}
C_{nm} &=&\delta_{nm}
     +\int d^{4}x\ \varphi_{n,i}^{\dag}(x)
        \left[i\alpha\gamma_{5}\right]_{ij}\varphi_{m,j}(x)\nonumber\\
     &&\ \ \ \ \ \ +\int d^{4}x\ \varphi_{n,i}^{\dag}(x)\frac{1}{2!}
        \left[i\alpha\gamma_{5}\right]_{ij}^{2}\varphi_{m,j}(x)\nonumber\\
     &&\ \ \ \ \ \ +\int d^{4}x\ \varphi_{n,i}^{\dag}(x)\frac{1}{3!}
        \left[i\alpha\gamma_{5}\right]_{ij}^{3}\varphi_{m,j}(x)
     +\cdot\cdot\cdot.\label{eq,29}
\end{eqnarray}
To proceed, we first derive the following identity by using the 
completeness relation given in Eq. (\ref{eq,28}):
\begin{eqnarray}
 \int d^{4}x\ \varphi_{n,i}^{\dag}(x)
        \left[i\alpha\gamma_{5}\right]_{ij}^{2}\varphi_{m,j}(x)
        \hspace{7cm}\nonumber\\
 =\int d^{4}x\int d^{4}y\ \varphi_{n,i}^{\dag}(x)
        \left[i\alpha\gamma_{5}\right]_{ij}\delta_{jk}\delta^{4}(x-y)
        \left[i\alpha\gamma_{5}\right]_{kl}\varphi_{m,l}(y)
        \hspace{0.28cm}\nonumber\\
 =\int d^{4}x\int d^{4}y\ \varphi_{n,i}^{\dag}(x)
        \left[i\alpha\gamma_{5}\right]_{ij}\varphi_{p,j}(x)
    \varphi_{p,k}^{\dag}(y)
        \left[i\alpha\gamma_{5}\right]_{kl}\varphi_{m,l}(y)
        \hspace{0cm}\nonumber\\
 =\tilde{C}_{np} \tilde{C}_{pm}=\tilde{C}^2_{nm},\hspace{7.25cm}
\end{eqnarray}
where
\begin{eqnarray}
   \tilde{C}_{nm}
   &=&\int d^{4}x\ \varphi_{n,i}^{\dag}(x)
     \left[i\alpha\gamma_{5}\right]_{ij}\varphi_{m,j}(x).
\end{eqnarray}
In a similar manner, we can prove the identity
\begin{eqnarray}
\int d^{4}x\ \varphi_{n,i}^{\dag}(x)
     \left[i\alpha\gamma_{5}\right]_{ij}^{N}\varphi_{m,j}(x)
     =\tilde{C}^N_{nm}.
     \end{eqnarray}
Therefore we can rewrite Eq. (\ref{eq,29}) as
\begin{eqnarray}
\int d^{4}x\ \varphi_{n,i}^{\dag}(x)
         \left(e^{i\alpha\gamma_{5}}\right)_{ij}
         \varphi_{m,j}(x)
         &=&
\delta_{nm}+\tilde{C}_{nm}+\frac{1}{2!}\tilde{C}^2_{nm}+
\frac{1}{3!}\tilde{C}^3_{nm}+\cdots~.
\end{eqnarray}
Then we use $\det A=\exp\{{\rm Tr}\ln A\}$ to obtain
\begin{eqnarray}
 J^{-1}
  &=&\det\left\{\int d^{4}x\ \varphi_{n,i}^{\dag}(x)
         \left(e^{i\alpha\gamma_{5}}\right)_{ij}
         \varphi_{m,j}(x)\right\}^{-2}\nonumber\hspace{3cm}\\
  &=&\exp\left\{-2\sum_{n}^{\infty}\int d^{4}x\ \varphi_{n,i}^{\dag}(x)
        \left[i\alpha\gamma_{5}\right]_{ij}\varphi_{n,j}(x)\right\}.
    \label{eq,30}
\end{eqnarray}
Note that in obtaining Eq. (\ref{eq,30}) we did not assume that the 
transformation parameter $\alpha$ is infinitesimal.

Next we use the quantity
\begin{eqnarray}
 \lim_{\Lambda\rightarrow\infty}e^{-(\lambda_{n}/\Lambda)^{2}}
\end{eqnarray}
as a regulator for the divergent summation, and we find
\begin{eqnarray}
 J^{-1}
   &=&\exp\left\{-2\lim_{\Lambda\rightarrow\infty}\sum_{n}^{\infty}
       \int d^{4}x\ \varphi_{n,i}^{\dag}(x)[i\alpha\gamma_{5}]_{ij}
       e^{-(\lambda_{n}/\Lambda)^{2}}\varphi_{n,j}(x)\right\}
       \nonumber\\
   &=&\exp\left\{-2i\lim_{\Lambda\rightarrow\infty}{\rm Tr}
       \int\frac{d^{4}k}{(2\pi)^{4}}\int d^{4}x\ e^{-ikx}
       \alpha\gamma_{5}e^{-(\not{D}/\Lambda)^{2}}
       e^{ikx}\right\}\nonumber\\
   &=&\exp\left\{-2i\lim_{\Lambda\rightarrow\infty}{\rm Tr}
       \int\frac{d^{4}k}{(2\pi)^{4}}\int d^{4}x\ \alpha\gamma_{5}\right.
       \nonumber\\
   &&\hspace{2cm}\times\left.\exp\left\{\frac{1}{\Lambda^{2}}
         \left[-(ik_{\mu}+D_{\mu})^{2}-\frac{i}{4}[\gamma^{\mu}\ 
         \gamma^{\nu}]F_{\mu\nu}\right] \right\}\right\}.
\end{eqnarray}
Since we are working  in the Euclidean space-time, we have the metric 
$g^{\mu\nu}={\rm diag}(-1,-1,-1,-1)$.
As is well known,\cite{fujikawa} \ the limiting procedure yields
\begin{eqnarray}
 J^{-1}=\exp\left\{-i\int d^{4}x\ 
         \frac{\alpha}{16\pi^{2}}{\rm Tr}
         \left[\epsilon^{\mu\nu\rho\sigma}
               F_{\mu\nu}F_{\rho\sigma}\right]\right\}.
               \label{eq,3}
\end{eqnarray}
In this way we can define anomalies of discrete symmetries.

\section{Anomalies of non-Abelian discrete family symmetries}
We start with the Lagrangian 
\begin{eqnarray}
 &&{\cal L}=i\bar{\psi}\ /\hspace{-0.3cm}{D}\psi-\frac{1}{2g^{2}_L}
          {\rm Tr}[F^{\mu\nu}(L)F_{\mu\nu}(L)]
          -\frac{1}{2g^{2}_R}
          {\rm Tr}[F^{\mu\nu}(R)F_{\mu\nu}(R)],\\
 &&D_{\mu}=\partial_{\mu}-iL_{\mu}^{a}T^{a}_LP_{L}
                       -iR_{\mu}^{b}T^{b}_RP_{R},
                       \label{lagrangian2}
\end{eqnarray}
which describes an $SU(N_L)\times SU(N_R)$ chiral Yang-Mills theory.
Here $L^{a}$ ($R^{a}$) are gauge bosons that couple to the left-handed 
(right-handed) fermions, and $P_{L}$ and $P_{R}$ are the projection 
operators on the left-handed and right-handed pieces, respectively.
Then we consider the non-abelian discrete chiral transformation
\begin{eqnarray}
 \psi_{i,\alpha^{'}}(x)\rightarrow\psi^{'}_{i,\alpha}=
   \left[e^{iXP_{L}+iYP_{R}}\right]_{ij,\alpha\beta}\psi_{j,\beta},
   \label{eq,18}
\end{eqnarray}
where 
$X$ and $Y$ are matrices that act on the family indices 
$\alpha,\beta=1-3$.
This transformation is a unitary transformation that does not mix the 
left-handed and right-handed fields.
Accordingly, we define the following phases:
\begin{eqnarray}
 &&e^{i\alpha_{L}}\equiv\det(e^{iX}),\hspace{1cm}
   e^{i\alpha_{R}}\equiv\det(e^{iY}),\label{eq,21}\\
 &&\alpha_{L}\equiv {\rm Tr}(X),\hspace{1.7cm}
   \alpha_{R}\equiv {\rm Tr}(Y).\label{eq,19}
\end{eqnarray}
In contrast to the previous case, we introduce two complete sets
of eigen-states:
\begin{eqnarray}
 &&/\hspace{-0.3cm}{D}^{\dag}\ /\hspace{-0.3cm}{D}\varphi_{n}
 =\lambda_{n}^{2}\varphi_{n},\hspace{2.7cm}
 /\hspace{-0.3cm}{D}\ /\hspace{-0.3cm}{D}^{\dag}\phi_{n}
                                  =\lambda_{n}^{2}\phi_{n},\\
 &&\psi_{i,\alpha}(x)=\sum_{n}a_{n}
 \varphi_{n,i,\alpha}(x),\hspace{1cm}
 \bar{\psi}_{i,\alpha}(x)=\sum_{n}\bar{b}_{n}
 \phi_{n,i,\alpha}^\dag(x).
\end{eqnarray}
Using these expressions, we then calculate the Jacobian for the 
transformation (\ref{eq,18}), where we denote the Jacobian for 
${\cal D}\psi$ by $j$ and that for ${\cal D}\bar{\psi}$ by $\bar{j}$, 
so that the total Jacobian $J$ is given by $\bar{j} j$.
After calculations similar to those in the abelian case, we find
\begin{eqnarray}
  j^{-1}&=&\left\{\det\int d^{4}x\ \varphi_{n,i,\alpha}^{\dag}(x)
              \left[e^{iXP_{L}+iYP_{R}}\right]_{ij,\alpha\beta}
              \varphi_{m,j,\beta}(x)\right\}^{-1}\nonumber\\
        &=&\exp\left\{-i\int d^{4}x\ \varphi_{n,i,\alpha}^{\dag}(x)
         \left[XP_{L}+YP_{R}\right]_{ij,\alpha\beta}
         \varphi_{m,j,\beta}(x)\right\}\nonumber\\
        &=&\exp\left\{-i\lim_{\Lambda\rightarrow\infty}{\rm Tr}
          \int\frac{d^{4}k}{(2\pi)^{4}}\int d^{4}x\ e^{-ikx}
          \left[XP_{L}+YP_{R}\right]
          e^{-(\not{D}^{\dag}\not{D}/\Lambda^{2})}e^{ikx}
          \right\},\label{eq,20}
\end{eqnarray}
and
\begin{eqnarray}
    \bar{j}^{-1}&=&
    \left\{\det\int d^{4}x\ \phi_{n,i,\alpha}^{\dag}(x)
              \left[e^{-iXP_{R}-iYP_{L}}\right]_{ij,\alpha\beta}
              \phi_{m,j,\beta}(x)\right\}^{-1}\nonumber\\
        &=&\exp\left\{i\int d^{4}x\ \phi_{n,i,\alpha}^{\dag}(x)
         \left[XP_{R}+YP_{L}\right]_{ij,\alpha\beta}
         \phi_{m,j,\beta}(x)\right\}\nonumber\\
        &=&\exp\left\{i\lim_{\Lambda\rightarrow\infty}{\rm Tr}
          \int\frac{d^{4}k}{(2\pi)^{4}}\int d^{4}x\ e^{-ikx}
          \left[XP_{R}+YP_{L}\right]
          e^{-(\not{D}\not{D}^{\dag}/\Lambda^{2})}e^{ikx}
          \right\},\label{eq,20b}
        \end{eqnarray}
where ${\rm Tr}$ stands for the trace over the spinor, family and Yang-Mills 
indices.
Carrying out the trace in the family space, we obtain
\begin{eqnarray}
  j^{-1}=\exp\left\{-i\lim_{\Lambda\rightarrow\infty}{\rm Tr}
          \int\frac{d^{4}k}{(2\pi)^{4}}\int d^{4}x\ e^{-ikx}
          \left[\alpha_{L}P_{L}+\alpha_{R}P_{R}\right]
          e^{-(\not{D}^{\dag}\not{D}/\Lambda^{2})}e^{ikx}\right\}
\end{eqnarray}
where use has been made of Eqs. (\ref{eq,21}) and (\ref{eq,19}).
In the same way, we obtain a similar expression for $\bar{j}^{-1}$.
The rest of the calculations are very similar to those in the previous 
case, and we finally obtain the total Jacobian,
\begin{eqnarray}
 J^{-1}&=&j^{-1} \bar{j}^{-1}\nonumber\\
  &=&\exp\left\{ i\int d^{4}x\frac{1}{32\pi^{2}}
       {\rm Tr}\epsilon^{\mu\nu\rho\sigma}\left[
         \alpha_{L}\ F_{\mu\nu}(L)F_{\rho\sigma}(L)
            -\alpha_{R}\ F_{\mu\nu}(R)F_{\rho\sigma}(R)\right]\right\},
       \label{eq,11}
\end{eqnarray}
where
\begin{eqnarray}
 &&F_{\mu\nu}(L)=\partial_{\mu}L_{\nu}-\partial_{\nu}L_{\mu}
                    -i[L_{\mu}\ L_{\nu}],\\
 &&F_{\mu\nu}(R)=\partial_{\mu}R_{\nu}-\partial_{\nu}R_{\mu}
                    -i[R_{\mu}\ R_{\nu}].
\end{eqnarray}
Here, $\alpha_{R}$ and $\alpha_{L}$ are the phases of the transformation 
matrices defined in Eq. (\ref{eq,21}), which need not be continuous nor 
infinitesimal.
These phases correspond to the abelian parts of the non-abelian discrete
family transformations.
Therefore, to calculate the anomaly of a non-abelian discrete family
symmetry, we only have to take into account its abelian parts.
We will consider specific examples in the later sections.

\section{Pontryagin index}
The expression 
\begin{eqnarray}
\int d^{4}x\ \frac{1}{32\pi^{2}}{\rm Tr}
 \left[\epsilon^{\mu\nu\rho\sigma}F_{\mu\nu}F_{\rho\sigma}\right]
 \end{eqnarray}
in Eq. (\ref{eq,11})
is called the Pontryagin index, which  is an integer $\nu$.
For the case  $\alpha_{L}\neq0$ and $\alpha_{R}=0$, for instance,
the Jacobian becomes
\begin{eqnarray}
 J^{-1}=\exp\{i\alpha_{L}\nu\}.
\end{eqnarray}
Since only the abelian parts of a non-abelian discrete family symmetry 
contribute to the anomalous Jacobian, the phase has the form
\begin{eqnarray}
\alpha_{L}=\sum_{I}~(2\pi/N_I)\sum_{i} q_{I,i}, 
\end{eqnarray}
where we have assumed that the abelian parts can be written as 
$\Pi_{I=1}Z_{N_I}$, and $ q_{I, i}$ represents the charge of $Z_{N_I}$.
Therefore, we obtain $J^{-1}=1$ if the relation
\begin{eqnarray}
 \sum_{i}q_{iI, }= r N_I\ \ \ (r=0,\pm1, \pm2\cdots)
\end{eqnarray}
is satisfied for each $I$.

\section{Anomaly cancellation  and gauge coupling unification}
In the previous sections we have seen that anomalies of discrete
symmetries can be defined as the anomalous Jacobian of the 
path-integral measure.
In this section, we study how an anomaly can be canceled by 
the Green-Schwarz (GS) mechanism.\cite{GS,4dim-GS,U(1)a} \ 
First, we review the GS mechanism for an anomalous $U(1)_{A}$ symmetry, 
and then we apply the GS mechanism to discrete symmetries.

\subsection{Green-Schwarz mechanism }
String theory when compactified on four dimensions usually contains
anomalous $U(1)$ local symmetries.
Consider a supersymmetric Yang-Mills theory based on a gauge group
${\cal G}\otimes U(1)_A$, where $U(1)_{A}$ is assumed to be anomalous.
The $U(1)_{A}$ gauge transformation is defined as
\begin{eqnarray}
 &&\Phi\rightarrow e^{-i\Lambda}\Phi,\\
 &&V_{A}\rightarrow V_{A}+i(\Lambda-\bar{\Lambda}),\hspace{1cm}
\end{eqnarray}
where $\Phi$ and $\Lambda$ are chiral supermultiplets and $V_{A}$ is
the vector supermultiplet of $U(1)_{A}$.
The anomaly for this transformation is calculated in Ref.~\cite{konishi}. 
Using the result of Ref.~\cite{konishi}, we find that the anomalous 
Jacobian for the $[{\cal G}]^2 \times U(1)_A$ anomaly, for instance, 
is given by
\begin{eqnarray}
J^{-1}=\exp\left\{-i{\cal A}\int d^4x
d^2\theta\ {\rm Tr}\left[\Lambda W^{a}W_{a}\right]_{F}\right\},
   \label{eq,8}
\end{eqnarray}
where ${\cal A}$ is the anomaly coefficient for 
$[{\cal G}]^2\times U(1)_A$ and $W$ is the chiral supermultiplet of 
the gauge supermultiplet corresponding to the gauge group ${\cal G}$.
This anomaly is canceled by the gauge kinetic term,
\begin{eqnarray}
 k\ {\rm Tr}\left[SW^{a}W_{a}\right]_{F},
\end{eqnarray}
if we correspondingly shift the dilaton supermultiplet $S$ as 
\begin{eqnarray}
 S\rightarrow S^{'}=S+i\frac{\cal A}{k}\Lambda,
 \label{eq,7}
\end{eqnarray}
where $k$ is the Kac-Moody level.
We must simultaneously modify the transformation property of the dilaton 
supermultiplet $S$ to restore the invariance of its K\"{a}hler potential.
At the quantum level, the K\"{a}hler potential is modified to include
$\delta_{GS}V_{A}$ in the logarithm:
\begin{eqnarray}
 K=\ln(S+\bar{S}-\delta_{GS}V_{A}).
 \label{kaehler}
\end{eqnarray}
Thus, the K\"{a}hler potential is invariant if the relation
\begin{eqnarray}
 \frac{\cal A}{k}=\delta_{GS}\label{eq,22}
\end{eqnarray}
is satisfied.

In the case ${\cal G}=SU(3)_{C}\otimes SU(2)_{L}\otimes U(1)_{Y}$,
there exist various possibilities for the anomaly:
$[SU(3)_{C}]^{2}\times U(1)_{A}\ $,
$\ [SU(2)_{L}]^{2}\times U(1)_{A}\ $,
$\ [U(1)_{Y}]^{2}\times U(1)_{A}\ $,$\ [U(1)_{A}]^{3}\ $ and
$\ [{\rm gravity}]^{2}\times U(1)_{A}$.
If we denote the respective anomaly coefficients by
${\cal A}_{3}$, ${\cal A}_{2}$, ${\cal A}_{1}$, ${\cal A}_{A}$,
and ${\cal A}_{G}$, then the anomaly cancellation conditions are 
given by
\begin{eqnarray}
 \frac{{\cal A}_{3}}{k_{3}}=\frac{{\cal A}_{2}}{k_{2}}=
 \frac{{\cal A}_{1}}{k_{1}}=\frac{{\cal A}_{A}}{k_{A}}=
 \frac{{\cal A}_{G}}{12}=\delta_{GS}.\label{eq,25}
\end{eqnarray}
Note that the Kac-Moody levels of a non-abelian group are positive 
integers, while there is no restriction on the Kac-Moody levels for an 
abelian group.

\subsection{Gauge coupling unification}
Next we consider the above discussion in terms of the component fields
to determine the relation of the Kac-Moody levels to the gauge coupling 
unification.
To this end, we define $S|_{\theta=\bar{\theta}=0}=\varphi+i\eta\ $ and
$\ \Lambda|_{\theta=\bar{\theta}=0}=\phi+i\xi$.
For the axion field $\eta$, the shift (\ref{eq,7}) corresponds to
\begin{eqnarray}
 \eta\rightarrow\eta^{'}=\eta+\phi\delta_{GS},
\end{eqnarray}
and the VEV of the dilaton field  $\varphi$ is merely the string
coupling, i.e.,
\begin{eqnarray}
 \langle\varphi\rangle=\frac{1}{g^{2}_{st}}.
\end{eqnarray}
Also, the gauge couplings $g_i$ are related to the string coupling 
according to  
\begin{eqnarray}
 \frac{k_{i}}{g^{2}_{st}}=\frac{1}{g^{2}_{i}}.
\end{eqnarray}
Therefore, the conditions for the gauge coupling unification of the 
SM gauge couplings can be written as
\begin{eqnarray}
 k_{3}g_{3}^{2}=k_{2}g_{2}^{2}=k_{1}g_{1}^{2}=g^{2}_{st}\label{eq,14}
\end{eqnarray}
at the string scale.
It is therefore clear that the anomaly cancellation conditions 
(\ref{eq,25}) have a non-trivial influence on the gauge coupling 
unification.

\subsection{The GS mechanism for discrete symmetries}
Here we extend the GS mechanism to the case of discrete symmetries.
Unlike Ref.~\cite{babu}, we do not assume that the discrete symmetry in 
question arises from a spontaneous breakdown of a continuous local 
symmetry.
We instead assume that the anomalous discrete symmetry at low energies 
is a remnant of an anomaly-free discrete symmetry, and that its low 
energy anomaly is cancelled by the GS mechanism at a more fundamental 
scale.
In superstring theory when compctified on a six-dimensional Calabi-Yau
manifold, for instance, there indeed exist certain non-abelian discrete 
symmetries.\cite{suematsu}

We consider the $Z_{N}$ transformation
\begin{eqnarray}
 &&\Phi\rightarrow e^{-i\alpha}\Phi,\label{eq,10}\\
 &&V\rightarrow V\hspace{1.7cm}\nonumber
 \label{susy-chiral-T}
\end{eqnarray} 
in a supersymmetric gauge theory in order to determine how to cancel 
the anomaly.
As in the previous case, the transformation parameter $\alpha$ is
discrete, i.e. $\alpha=\frac{2\pi}{N}$, and $V$ is the vector 
supermultiplet of the gauge group.
The anomaly of this transformation has the same form as (\ref{eq,8}):
\begin{eqnarray}
 J^{-1}=\exp\left\{-i{\cal A}\int d^4x d^2\theta\ {\rm Tr}
                  \left[\alpha W^{a}W_{a}\right]_{F}\right\},
\end{eqnarray}
which is also canceled by the gauge kinetic term.
In this case, however, the dilaton supermultiplet must be shifted by
only a constant amount $\alpha$, i.e.,
\begin{eqnarray}
 S\rightarrow S^{'}=S+i\frac{\cal A}{k}\alpha,
\end{eqnarray}
for the cancellation mechanism to work.
This means that because $\alpha$ is a constant, independent of $x$ and 
$\theta$, only the imaginary part of the scalar component of $S$, which 
is the axion field, should be shifted.
Note that the K\"{a}hler potential (\ref{kaehler}) does not depend on
$\delta_{GS}$ because the vector supermultiplet does not change under 
the transformation Eq. (\ref{eq,10}).
Therefore, the anomaly cancellation conditions for the SM gauge group
are
\begin{eqnarray}
  \frac{{\cal A}_{3}}{k_{3}}=\frac{{\cal A}_{2}}{k_{2}}=
  \frac{{\cal A}_{1}}{k_{1}}=
  \frac{{\cal A}_{G}}{12},
\end{eqnarray}
where ${\cal A}_{3}$, ${\cal A}_{2}$, ${\cal A}_{1}$ and ${\cal A}_{G}$
are the anomaly coefficients of the anomalies
$[SU(3)_{C}]^{2}\times Z_{N}\ $, $\ [SU(2)_{L}]^{2}\times Z_{N}\ $,
$\ [U(1)_{Y}]^{2}\times Z_{N}\ $
and 
$\ [{\rm gravity}]^{2}\times Z_{N}$, respectively.
Because the $\ [U(1)_{Y}]^{2}\times Z_{N}\ $ anomaly does not yield
useful constraints on the low-energy effective theory, and we cannot 
calculate ${\cal A}_{G}$ for the low-energy effective theory, we do not 
consider them when studying models in the next section.
However, massive Majorana fields can contribute to ${\cal A}_{2}$ and 
${\cal A}_{3}$ for even $N$, because  Majorana masses are allowed by the 
discrete symmetry if the Majorana fields belong to real representations
of $SU(2)_L$ and $SU(3)_C$.\cite{babu} \ 
Taking into account the contributions from the massive fields, we arrive 
at the anomaly cancellation conditions\cite{babu}
\begin{eqnarray}
 \frac{{\cal A}_{3}+\frac{pN}{2}}{k_{3}}
   =\frac{{\cal A}_{2}+\frac{qN}{2}}{k_{2}},\label{eq,27}
\end{eqnarray}
with integer $p$ and $q$, where $pN/2$ and $qN/2$ take into account the 
possible contributions from the heavy fields.

In this section, we have applied the GS mechanism to an abelian
discrete symmetry.
In \S7, we consider anomaly cancellation of a non-abelian discrete 
family symmetry.
As we have seen in \S3, however, only the abelian parts contribute
to the anomaly, even if we consider a non-abelian discrete
family symmetry.
Hence, Eq. (\ref{eq,27}) can also be applied to the case of a non-abelian
discrete family symmetries.

\section{Anomalies of accidental $Z_{N}$ symmetries}
As an example of anomaly cancellation for discrete symmetries, let us
consider the baryon and lepton number symmetries in the MSSM with 
R-parity, in which the see-saw mechanism is implemented to generate 
neutrino masses.
The baryon number $U(1)_B$ is conserved at the classical level, 
while the lepton number $U(1)_L$ is not conserved because of the
Majorana masses of  the right-handed neutrinos.
However, its abelian discrete subgroups, $(Z_{N})_{L}$ with even $N$, 
are intact at the classical level.
In the following discussion, we first investigate anomalies and their
GS cancellation conditions for abelian discrete subgroups  
$(Z_{M})_{B}$ of $U(1)_B$ and  $(Z_{N})_{L}$.
Then we study how the GS cancellation conditions influence gauge 
coupling unification.
The $(Z_{N})_{L}$ and $(Z_{M})_{B}$ charges of the supermultiplets
of the MSSM are given in Table \ref{table1}.
\begin{table}[h]
\begin{center}
\caption{The $(Z_{N})_{L}$ and $(Z_{N})_{B}$ charges.
$N$ is even and $a,b,c,\cdots p=0,1,2\cdots$.
}
\begin{tabular}{|c|c|c|c|c|c|c|c|c|}\hline\hline
    & $Q$ & $U^{c}$ & $D^{c}$ & $L$ &
                 $E^{c}$ & $\nu_{R}$ & $H^{u}$ & $H^{d}$\\ \hline
 $(Z_{N})_{L}$ & $aN$ & $bN$ & $cN$ & $\frac{N}{2}+dN$ &
            $\frac{N}{2}+eN$ & $\frac{N}{2}+fN $ & $gN$ & $hN$\\ \hline
 $(Z_{M})_{B}$ & $B+iM$ & $-B+jM$ & $-B+kM$ &
                           $lM$ & $mM$ & $nM$ & $oM$ & $pM$\\ \hline
\end{tabular}\label{table1}
\end{center}
\end{table}

The anomaly coefficients for $(Z_{N})_{L}$ are found to be
\begin{eqnarray}
 &&2{\cal A}_{3}=\frac{N}{2}[12a+6b+6c],\\
 &&2{\cal A}_{2}=\frac{N}{2}[18a+3+6d+2g+2h].
\end{eqnarray}
Therefore, the GS cancellation conditions become
\begin{eqnarray}
 \frac{k_{3}}{k_{2}}=\frac{12a+6b+6c}{18a+3+6d+2g+2h}
         =\frac{\rm even}{\rm odd}.
 \label{eq,31}
\end{eqnarray}
Note that $k_3=k_2=1$ is {\it not} a solution.
In a similar way, we can calculate the anomaly coefficients for
$(Z_{M})_{B}$, and we find
\begin{eqnarray}
 \frac{k_{3}}{k_{2}}
   =\frac{9B+(9i+3\ell+o+p)M}{(6i+3j+3k)M}
   =\frac{{\rm even}\ {\bf or}\ {\rm odd}}
         {{\rm even}\ {\bf or}\ {\rm odd}}.\label{eq,32}
\end{eqnarray}
As it is believed to be difficult to build realistic models with
higher Kac-Moody levels  in string theory, we seek solutions to 
(\ref{eq,31}) and (\ref{eq,32}) with lower levels.
The solution with the lowest levels that satisfy the conditions
(\ref{eq,31}) and (\ref{eq,32}) simultaneously is $k_3=2,k_2=1$, 
which yields the gauge coupling unification conditions
\begin{eqnarray}
 2g_3^{2}=g_{2}^{2}=k_{1}g_{1}^{2}=g_{st}^{2},
\end{eqnarray}
where $k_1$ is arbitrary.
Figure \ref{fig,1} plots the ratios $g_2^2/g_1^2$ (upper curves)
and $g_2^2/g_3^2$ (lower curves) as functions of the energy scale.
The solid curves correspond to the case with one pair of $SU(2)_L$ 
doublet Higgs supermultiplets, the dotted curves to the case with two 
pairs, and the dashed curves to the case with three pairs.
(We denote the number of the Higgs pairs by $H_{\rm higgs}$.)
We see from Figure \ref{fig,1} that the ratio $g_2^2/g_3^2$ with 
$H_{\rm higgs}=3$ becomes close to $2$ at the Planck scale, 
$M_{PL}=1.2\times 10^{18}$ GeV.
For $H_{\rm higgs}=1$ and $2$ there is no chance for the ratio to become 
close to $2$ below or near  $M_{PL}$.
In Figure \ref{fig,2} we plot the running of 
$(\alpha_1 k_1)^{-1},(\alpha_2 k_2)^{-1}$ and
$(\alpha_3 k_3)^{-1}$ with $k_{3}=2,k_{2}=1$ and $k_{1}=2.25$
in the case $H_{\rm higgs}=3$.
As we have seen above, the GS cancellation conditions of anomalies have 
a nontrivial influence on the gauge coupling unification, and hence the 
number of Higgs supermultiplets.
\begin{figure}[p]
 \begin{center}
  \includegraphics[width=13cm,clip]{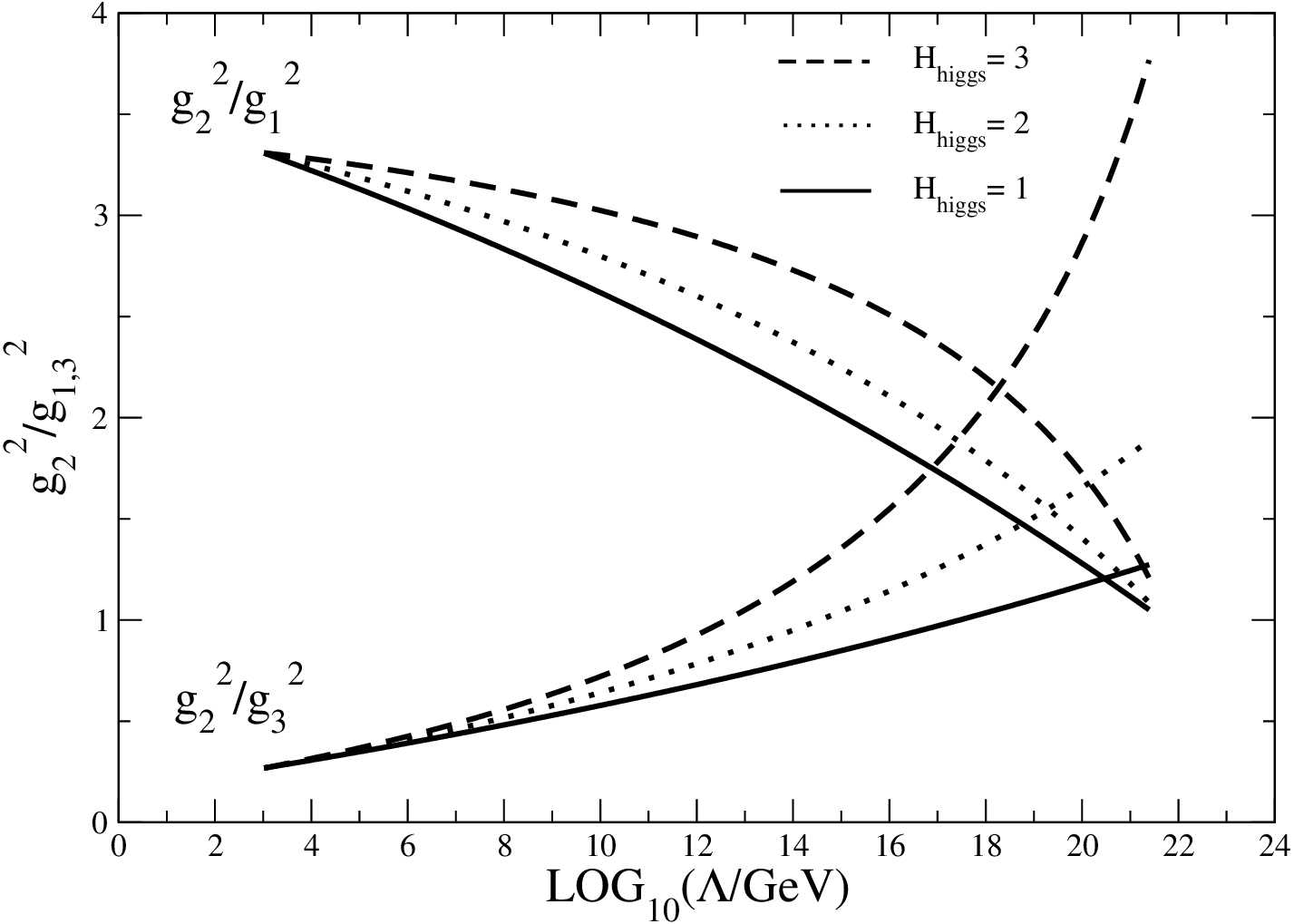}
  \caption{\footnotesize
The ratios $g_2^2/g_1^2$ (upper curves) and $g_2^2/g_3^2$ (lower curves)
as functions of the energy scale.
The solid curves correspond to the MSSM+$\nu_{R}$ case, the dotted
curves to the case with $H_{\rm higgs}=2$, and the dashed curves to the 
case with $H_{\rm higgs}=3$.}
  \label{fig,1}
  \vspace{0.5cm}
  \includegraphics[width=13cm,clip]{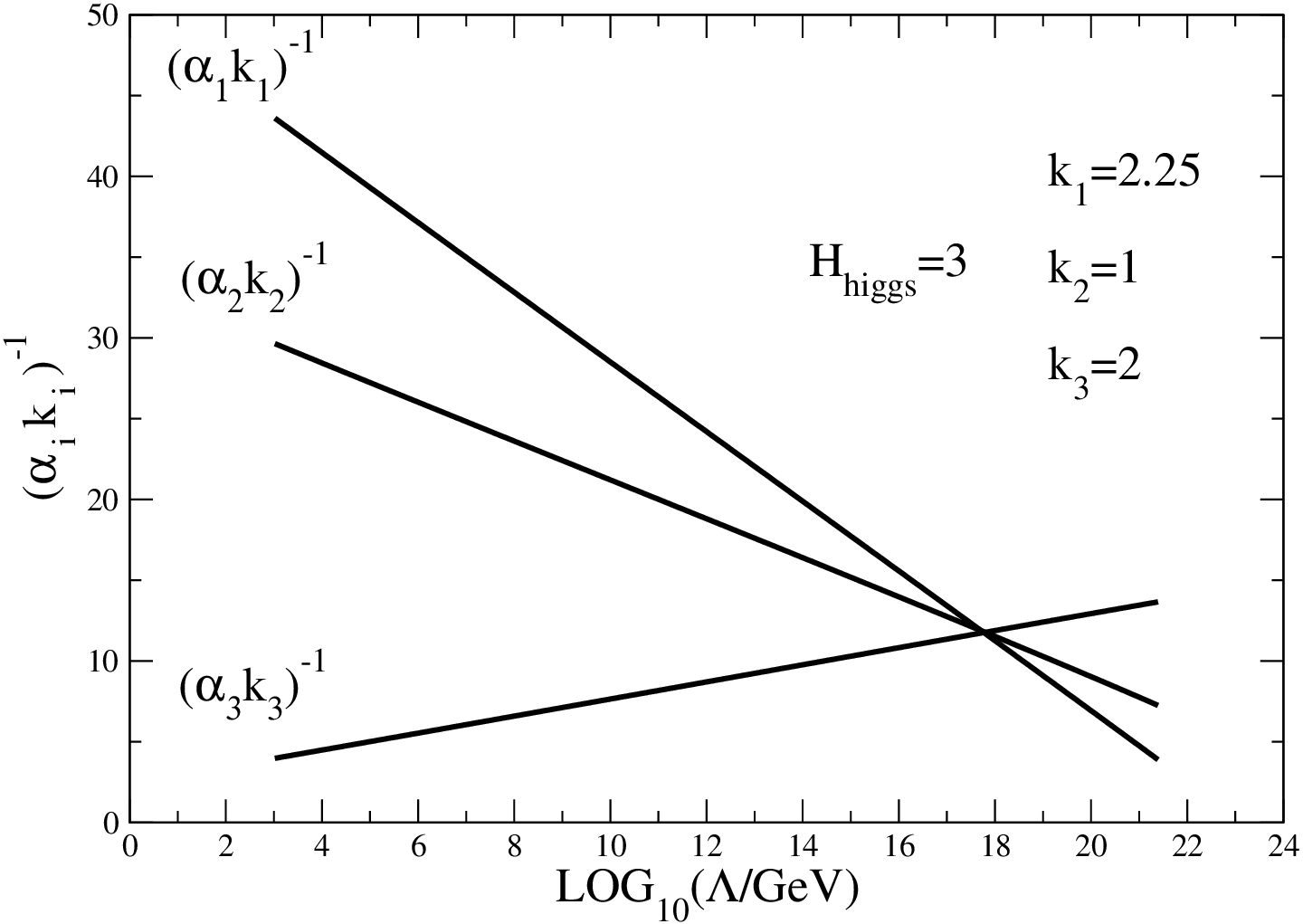}
  \caption{\footnotesize
The running of $(\alpha_1 k_1)^{-1}$, $(\alpha_2 k_2)^{-1}$
and $(\alpha_3 k_3)^{-1}$ with $k_{3}=2, k_{2}=1$ and $ k_{1}=2.25$
in the case with $H_{\rm higgs}=3$.
The unification scale is $10^{18}$ GeV.}
  \label{fig,2}
 \end{center}
\end{figure}

\section{Models}
Recently, a number of models with a non-abelian discrete family symmetry
have been proposed.\cite{A4,S3,D4,Q6,D7,D5,AS4} \ 
However, if only the SM Higgs or the MSSM Higgs are present within the 
framework of renormalizable models, any low-energy non-abelian family 
symmetry should be hardly broken to be consistent with experimental 
observations.
That is, if a non-abelian discrete family symmetry should be at most
softly broken, we need several pairs of  $SU(2)_L$ doublet Higgs fields.
This implies that the conditions of the ordinary unification of gauge
couplings, i.e. $k_2=k_3=k_1(3/5)$, cannot be satisfied if we require
a low-energy non-abelian discrete family symmetry.
Fortunately, however, as we have seen, there is a possibility to satisfy
the gauge coupling unification conditions with non-minimal Higgs content
if the Kac-Moody levels assume non-trivial values.
These Kac-Moody levels also play an important role in the anomaly 
cancellation (GS mechanism).
In the following subsections, we calculate the anomalies of non-abelian
discrete family symmetries for recently proposed models and
investigate the gauge coupling unification conditions.

\subsection{$D_{7}$ model}
Let us first calculate the anomaly of the supersymmetric $D_{7}$
model.\cite{D7,D5} \ 
$D_{7}$ has fourteen elements and five irreducible representations
$({\bf 1,1^{'},2,2^{'},2^{''}})$.
This model uses the complex representation,\cite{complex} \ and thus 
the character table and the two-dimensional representation matrices 
of ${\bf 2}$ are as follows:
\begin{table}[h]
\begin{center}
\caption{Character table of $D_{7}$.\hspace{2.3cm}}
\begin{tabular}{|c|c|c|c|c|c|c|c|}\hline\hline
 class & n & h & $\chi_{1}$ & $\chi_{1^{'}}$ & $\chi_{2}$
                    & $\chi_{2^{'}}$ & $\chi_{2^{''}}$ \\ \hline
 $C_{1}$ & 1 & 1 & 1 & 1 & 2 & 2 & 2\\ \hline
 $C_{2}$ & 7 & 2 & 1 & $-1$ & 0 & 0 & 0\\ \hline
 $C_{3}$ & 2 & 7 & 1 & 1 & $a_{1}$ & $a_{2}$ & $a_{3}$\\ \hline
 $C_{4}$ & 2 & 7 & 1 & 1 & $a_{2}$ & $a_{3}$ & $a_{1}$\\ \hline
 $C_{5}$ & 2 & 7 & 1 & 1 & $a_{3}$ & $a_{1}$ & $a_{2}$\\ \hline
\end{tabular}\ \ $a_{k}=2\cos(\frac{2\pi}{7}k)$
\label{table2}
\end{center}
\end{table}
\begin{eqnarray}
 &&C_{1}:
  \left(\begin{array}{cc}
    1 & 0 \\
    0 & 1
  \end{array}\right), \nonumber\\
 &&C_{2}:
  \left(\begin{array}{cc}
    0 & \omega^{k} \\
    \omega^{7-k} & 0
  \end{array}\right)\ \ k=0-6\ \ \ \omega=\exp(2\pi i/7),\nonumber\\
 &&C_{3}:
  \left(\begin{array}{cc}
    \omega^{6} & 0 \\
    0 & \omega
  \end{array}\right),
  \left(\begin{array}{cc}
    \omega & 0 \\
    0 & \omega^{6}
  \end{array}\right),\nonumber\\
 &&C_{4}:
  \left(\begin{array}{cc}
    \omega^{5} & 0 \\
    0 & \omega^{2}
  \end{array}\right),
  \left(\begin{array}{cc}
    \omega^{2} & 0 \\
    0 & \omega^{5}
  \end{array}\right),\nonumber\\
 &&C_{3}:
  \left(\begin{array}{cc}
    \omega^{4} & 0 \\
    0 & \omega^{3}
  \end{array}\right),
  \left(\begin{array}{cc}
    \omega^{3} & 0 \\
    0 & \omega^{4}
  \end{array}\right).\label{eq,33}
\end{eqnarray}
The representation matrices for ${\bf 2^{'}}$ and ${\bf 2^{''}}$ are
obtained from the cyclic rotation of $C_{3}$, $C_{4}$ and $C_{5}$.
$D_{7}$ has five kinds of transformation properties, corresponding to
five classes.
However, the transformation of $C_{1}$ is the identity, and there is 
no difference among $C_{3}-C_5$ when we calculate the anomaly.
Hence we consider only $C_{2}$ and $C_{3}$.
Under $C_{2}$ and $C_{3}$, the irreducible representations transform as
\begin{eqnarray}
 C_{2}\hspace{5.2cm}C_{3}\hspace{2.7cm}\nonumber\\ \nonumber\\
   \begin{array}{llllll}
     {\bf 2} &  & \rightarrow \left(\begin{array}{cc}
                                0 & \omega^{k} \\
                                \omega^{7-k} & 0
                              \end{array}\right){\bf 2}
       & & & \rightarrow \left(\begin{array}{cc}
                               \omega^{6} & 0 \\
                               0 & \omega^{1}
                         \end{array}\right),
                         \left(\begin{array}{cc}
                               \omega^{1} & 0 \\
                               0 & \omega^{6}
                         \end{array}\right){\bf 2},\\

     {\bf 2}^{'} &  & \rightarrow \left(\begin{array}{cc}
                                        0 & \omega^{k} \\
                                        \omega^{7-k} & 0
                                  \end{array}\right){\bf 2}^{'}
       & & & \rightarrow \left(\begin{array}{cc}
                               \omega^{5} & 0 \\
                               0 & \omega^{2}
                         \end{array}\right),
                         \left(\begin{array}{cc}
                               \omega^{2} & 0 \\
                               0 & \omega^{5}
                         \end{array}\right){\bf 2}^{'},\\

     {\bf 2}^{''} &  & \rightarrow \left(\begin{array}{cc}
                                         0 & \omega^{k} \\
                                         \omega^{7-k} & 0
                                   \end{array}\right){\bf 2}^{''}
       & & & \rightarrow \left(\begin{array}{cc}
                               \omega^{4} & 0 \\
                               0 & \omega^{3}
                         \end{array}\right),
                         \left(\begin{array}{cc}
                               \omega^{3} & 0 \\
                               0 & \omega^{4}
                         \end{array}\right){\bf 2}^{''},\\

     {\bf 1} &  & \rightarrow{\bf 1} & & & \rightarrow{\bf 1},\\

     {\bf 1}^{'} &  & \rightarrow e^{i\frac{2\pi}{2}1}\ {\bf 1}^{'} 
                   & & & \rightarrow {\bf 1}^{'},
   \end{array} \label{eq,15}
\end{eqnarray}
where $k=1-7$ and $\omega=\exp(2\pi i/7)$.
As mentioned in \S3, even if we calculate the anomaly of a
non-abelian discrete family symmetry, only the abelian parts contribute
to the anomaly.
From Eqs. (\ref{eq,33}) and (\ref{eq,15}), it is clear that the abelian
parts of the $C_{2}$ and $C_{3}$ transformations are $Z_{2}$ and $Z_{7}$,
respectively.

The authors of Refs.~\cite{D7} and \cite{D5} introduce into this model 
the $SU(2)_{L}$ triplet extra Higgs supermultiplets $\xi_{i}$, which have 
lepton numbers, and assume that all Higgs supermultiplets have generation
as well as fermions.
The assignment of the $D_{7}$ representations for each matter
supermultiplet is presented in Table \ref{table3}.
\footnote{
In Ref.~\cite{D7}, the assignment for leptons is not specified.
Therefore we employ the assignment for those supermultiplets given
in Ref.~\cite{D5}.}
\begin{table}[h]
\begin{center}
\caption{$D_{7}$ assignment of the matter supermultiplets.}
\begin{tabular}{|c|c|c|c|c|c|c|c|c|c|}\hline\hline
    & $Q_{1,2}\ D^{c}_{1,2}$ & $U^{c}_{1,2}$ & $L_{2,3}\ E^{c}_{2,3}$ 
    & $Q_{3}\ D^{c}_{3}\ U^{c}_{3}\ L_{1}\ E^{c}_{1}$
    & $H^{d}_{1,2}$ & $H^{u}_{1,2}$ & $\xi_{2,3}$
    & $H^{d}_{3}\ H^{u}_{3}\ \xi_{1}$ \\ \hline
  $D_{7}$ & 2 & $2^{'}$ & $2^{''}$ & 1 
              & 2 & $2^{''}$ & $2^{''}$ & 1 \\ \hline
\end{tabular}
\label{table3}
\end{center}
\end{table}

\subsubsection{Computation of anomaly coefficients}
Here we compute the anomaly coefficients for the $C_{3}(Z_{7})$ 
transformations.
These transformations have no anomaly, because the determinants of all
$C_{3}$ transformation matrices in Eq. (\ref{eq,15}) are equal to $1$.

Next we compute the anomaly coefficients for the $C_{2}(Z_{2})$
transformations.
The first and second generations of $QD^{c}$,$U^{c}$ and $H^{u}H^{d}$ are
assigned to ${\bf 2,2^{'}}$ and ${\bf 2^{''}}$, respectively, and the
third generation is assigned to ${\bf 1}$.
Hence, these supermutiplets transform as
\begin{eqnarray}
 \Psi_{\alpha=1-3}\rightarrow
   \left(\begin{array}{ccc}
     0 & \omega^{k} & 0 \\
     \omega^{7-k} & 0 & 0 \\
     0 & 0 & 1
   \end{array}\right)_{\alpha\beta}
 \Psi_{\beta=1-3}.
\end{eqnarray}
The determinant of this matrix is $\exp(2\pi i/2)=-1$, and therefore 
these supermultiplets contribute $\frac{2\pi}{2}1$ anomaly coefficients.
Here we omit the $Z_{N}$ factor, that is $\frac{2\pi}{N}$, and hence 
the contributions of these supermultiplets are equal to $1$.
The second and third generations of $L,E^{c}$ and $\xi$ are assigned to 
${\bf 2^{''}}$, and the first generation is assigned to ${\bf 1}$.
Hence, these supermultiplets transform as
\begin{eqnarray}
  \Phi_{\alpha=1-3}\rightarrow
   \left(\begin{array}{ccc}
     1 & 0 & 0 \\
     0 & 0 & \omega^{k} \\
     0 & \omega^{7-k} & 0
   \end{array}\right)_{\alpha\beta}
 \Phi_{\beta=1-3},
\end{eqnarray}
and contribute $1$ as well.
Using these facts, we can compute the anomaly coefficients and find
\begin{eqnarray}
 &&2{\cal A}_{3}=[1\cdot2+1+1]=4\ ({\rm mod}\ 2), \\
 &&2{\cal A}_{2}=[1\cdot3+1+1+1]+1\cdot4\cdot2=14\ ({\rm mod}\ 2).
 \label{eq,34}
\end{eqnarray}
Here, we define ${\cal A}_{3}$ and ${\cal A}_{2}$ as the anomaly
coefficients of $[SU(3)_{C}]^{2}\times Z_{2}$ and
$[SU(2)_{L}]^{2}\times Z_{2}$, respectively.
The last term in Eq. (\ref{eq,34}) is the contribution from the
$SU(2)_{L}$ triplet Higgs supermultiplets.
As we have seen in \S4, these coefficients do not contribute an 
anomaly.
Therefore this model is anomaly-free.

\subsection{$A_{4}$ model}
As the second example, we calculate the anomaly of the supersymmetric 
$A_{4}$ model.\cite{A4} \ 
$A_{4}$ has twelve elements and four irreducible representations
$({\bf 1,1^{'},1^{''},3})$.
This model also uses the complex representation, and thus the character
table and the three-dimensional representation matrices of
${\bf 3}$ are written as follows:
\begin{table}[h]
\begin{center}
\caption{Character table of $A_{4}$.\hspace{1.5cm}}
\begin{tabular}{|c|c|c|c|c|c|c|}\hline\hline
 class & n & h & $\chi_{1}$ & $\chi_{1'}$ 
                      & $\chi_{1''}$ & $\chi_{3}$ \\ \hline
 $C_{1}$ & 1 & 1 & 1 & 1 & 1 & 3 \\ \hline
 $C_{2}$ & 4 & 3 & 1 & $\omega$ & $\omega^{2}$ & 0 \\ \hline
 $C_{3}$ & 4 & 3 & 1 & $\omega^{2}$ & $\omega$ & 0 \\ \hline
 $C_{4}$ & 3 & 2 & 1 & 1 & 1 & $-1$ \\ \hline
\end{tabular}\ \ $\omega=e^{i\frac{2\pi}{3}}$
\label{table4}
\end{center}
\end{table}
\begin{eqnarray}
 &&C_{1}:
  \left(\begin{array}{ccc}
    1 & 0 & 0\\
    0 & 1 & 0\\
    0 & 0 & 1
  \end{array}\right), \nonumber\\
 &&C_{2}:
  \left(\begin{array}{ccc}
    0 & 0 & 1\\
    1 & 0 & 0\\
    0 & 1 & 0
  \end{array}\right),
  \left(\begin{array}{ccc}
    0 & 0 & 1\\
   -1 & 0 & 0\\
    0 &-1 & 0
  \end{array}\right),
  \left(\begin{array}{ccc}
    0 & 0 &-1\\
   -1 & 0 & 0\\
    0 & 1 & 0
  \end{array}\right),
  \left(\begin{array}{ccc}
    0 & 0 &-1\\
    1 & 0 & 0\\
    0 &-1 & 0
  \end{array}\right),\nonumber\\
 &&C_{3}:
  \left(\begin{array}{ccc}
    0 & 1 & 0\\
    0 & 0 & 1\\
    1 & 0 & 0
  \end{array}\right),
  \left(\begin{array}{ccc}
    0 & 1 & 0\\
    0 & 0 &-1\\
   -1 & 0 & 0
  \end{array}\right),
  \left(\begin{array}{ccc}
    0 &-1 & 0\\
    0 & 0 & 1\\
   -1 & 0 & 0
  \end{array}\right),
  \left(\begin{array}{ccc}
    0 &-1 & 0\\
    0 & 0 &-1\\
    1 & 0 & 0
  \end{array}\right),\nonumber\\
 &&C_{4}:
  \left(\begin{array}{ccc}
    1 & 0 & 0\\
    0 &-1 & 0\\
    0 & 0 &-1
  \end{array}\right),
  \left(\begin{array}{ccc}
   -1 & 0 & 0\\
    0 & 1 & 0\\
    0 & 0 &-1
  \end{array}\right),
  \left(\begin{array}{ccc}
   -1 & 0 & 0\\
    0 &-1 & 0\\
    0 & 0 & 1
  \end{array}\right).
\end{eqnarray}
$A_{4}$ has four kinds of transformation properties, corresponding to
four classes.
However, there is no difference between $C_{2}$ and $C_{3}$ when
we calculate the anomaly.
The abelian parts of $C_{3}$ and $C_{4}$ are $Z_{3}$ and $Z_{2}$, 
respectively.

The assignment of the $A_{4}$ representations for the matter
supermultiplets is given in Table \ref{table5}.
In addition, the authors of Ref.~\cite{A4} introduce the extra leptons, 
quarks and Higgs supermultiplets listed in Table \ref{table6}.
\begin{table}[h]
\begin{center}
\caption{$A_{4}$ assignment of the matter supermultiplets.}
\begin{tabular}{|c|c|c|c|c|c|}\hline\hline
    & $Q_{1,2,3}\ L_{1,2,3}$ 
    & $U^{c}_{1}\ D^{c}_{1}\ E^{c}_{1}$ 
    & $U^{c}_{2}\ D^{c}_{2}\ E^{c}_{2}$ 
    & $U^{c}_{3}\ D^{c}_{3}\ E^{c}_{3}$ 
    & $H^{u}\ H^{d}$ \\ \hline
 $A_{4}$ & 3 & $1$ & $1^{'}$ & $1^{''}$ & 1 \\ \hline
\end{tabular}
\label{table5}
\end{center}
\end{table}
\begin{table}[h]
\begin{center}
\caption{$A_{4}$ and $SU(3)_{C}$ assignments of the extra supermultiplets.}
\begin{tabular}{|c|c|c|c|c|c|}\hline\hline
    & $u_{1,2,3}\ u^{c}_{1,2,3}$ 
    & $d_{1,2,3}\ d^{c}_{1,2,3}$ 
    & $e_{1,2,3}\ e^{c}_{1,2,3}$ 
    & $N^{c}_{1,2,3}$ 
    & $\chi_{1,2,3}$ \\ \hline
 $A_{4}$ & 3 & 3 & 3 & 3 & 3 \\ \hline
 $SU(3)_{C}$ & 3 & 3 & 1 & 1 & 1 \\ \hline
\end{tabular}
\label{table6}
\end{center}
\end{table}

\subsubsection{Computation of anomaly coefficients}
We next compute the anomaly coefficients in the same way as in the 
case of $D_{7}$.
For the $C_{3}(C_2)$ transformation, the supermultiplets that are 
assigned to {\bf 3} and {\bf 1} do not contribute to anomaly, because 
the determinants of these transformation matrices are equal to 1. 
Therefore, only ${\bf 1^{'}}$ and ${\bf 1^{''}}$ contribute to
the anomaly coefficients.
As we can see from Table \ref{table5}, the three generations of
the right-handed quark and lepton are assigned to
${\bf 1},{\bf 1'{'}}$ and ${\bf 1^{'''}}$, respectively, and so they 
transform as
\begin{eqnarray}
 &&1{\rm st}\hspace{0.5cm}\ddag\ \ {\bf 1}\rightarrow {\bf 1},\\
 &&2{\rm nd}\hspace{0.38cm}\ddag\ \ {\bf 1^{'}}
           \rightarrow e^{i\frac{2\pi}{3}2}{\bf 1^{'}},\\
 &&3{\rm rd}\hspace{0.42cm}\ddag\ \ {\bf 1^{''}}
           \rightarrow e^{i\frac{2\pi}{3}1}{\bf 1^{''}}.
\end{eqnarray}
Therefore, these representations do not contribute to the anomaly
when we take into account all generations.

On the other hand, there is no anomaly for the $C_{4}$ transformation,
because all singlets do not transform, and the determinants of all
transformation matrices of {\bf 3} are equal to 1.
Therefore this model is anomaly-free.

\subsection{$Q_{6}$ model}
We finally calculate the anomaly of the supersymmetric $Q_{6}$
model.\cite{Q6} \ 
All elements of $Q_{6}$ are constructed from combinations of
\begin{eqnarray}
 A_{Q_{6}}=
   \left(\begin{array}{cc}
      \cos\ \phi_{6} & \sin\ \phi_{6} \\
     -\sin\ \phi_{6} & \cos\ \phi_{6}
   \end{array}\right)_{\phi_{6}=\frac{2\pi}{6}},\hspace{1cm}
 B_{Q}=
   \left(\begin{array}{cc}
      i & 0 \\
      0 & -i
   \end{array}\right),\label{eq,12}
\end{eqnarray}
as follows:
\begin{eqnarray}
 {\cal G}=\{E,A_{Q_{6}},(A_{Q_{6}})^{2},\cdots,(A_{Q_{6}})^{5},
            B_{Q},A_{Q_{6}}B_{Q},(A_{Q_{6}})^{2}B_{Q},\cdots,
            (A_{Q_{6}})^{5}B_{Q}\}.
\end{eqnarray}
Here, $E$ is the identity element.
$Q_{6}$ has six irreducible representations, two doublets and
four singlets:
\begin{eqnarray}
 {\rm doublet}&\ddag&{\bf 2}\ \ \ {\bf 2}^{'},\\
 {\rm singlet}&\ddag&{\bf 1}\ \ \ {\bf 1}^{'}\ \ \ 
 {\bf 1}^{''}\ \ \ {\bf 1}^{'''}.
\end{eqnarray}
The transformation properties of each irreducible representation
are characterized by $A_{Q_{6}}$ and $B_{Q}$.
For example, they can be written as follows:
\begin{eqnarray}
 A_{Q_{6}}\hspace{4cm}B_{Q}\hspace{1.3cm}\nonumber\\ \nonumber\\
  \begin{array}{llllll}
    {\bf 2} & & \rightarrow \left(\begin{array}{cc}
                               1/2 & \sqrt{3}/2 \\
                               -\sqrt{3}/2 & 1/2
                          \end{array}\right){\bf 2}
      & & & \rightarrow \left(\begin{array}{cc}
                             i & 0\\
                             0 & -i
                        \end{array}\right){\bf 2},\\

    {\bf 2}^{'} & & \rightarrow \left(\begin{array}{cc}
                                     -1/2 & \sqrt{3}/2 \\
                                     -\sqrt{3}/2 & -1/2
                                \end{array}\right){\bf 2}^{'}
      & & & \rightarrow \left(\begin{array}{cc}
                             1 & 0\\
                             0 & -1
                        \end{array}\right){\bf 2}^{'},\\

    {\bf 1} & & \rightarrow{\bf 1} & & & \rightarrow{\bf 1},\\

    {\bf 1}^{'} & & \rightarrow{\bf 1}^{'} 
             & & & \rightarrow e^{i\frac{2\pi}{4}2}\ {\bf 1},\\

    {\bf 1}^{''} & & \rightarrow e^{i\frac{2\pi}{6}3}\ {\bf 1}^{''}
             & & & \rightarrow e^{i\frac{2\pi}{4}3}\ {\bf 1}^{''},\\

    {\bf 1}^{'''} & &\rightarrow e^{i\frac{2\pi}{6}3}\ {\bf 1}^{'''}
             & & &\rightarrow e^{i\frac{2\pi}{4}1}\ {\bf 1}^{'''}.
  \end{array}
\end{eqnarray}
It is clear that the abelian parts of the $A_{Q_{6}}$ and $B_{Q}$ 
transformations are equal to $Z_{6}$ and $Z_{4}$, respectively, because
we have
\begin{eqnarray}
 (A_{Q_{6}})^{6}=E\hspace{1cm}(B_{Q})^{4}=E.
\end{eqnarray}

The assignment of the $Q_{6}$ representations of the matter
supermultiplets is given in Table \ref{table7}.
\begin{table}[h]
\begin{center}
\caption{$Q_{6}$ assignment of the matter supermultiplets.}
\begin{tabular}{|c|c|c|c|c|c|c|}\hline\hline
          & $Q_{1,2}\ L_{1,2}$ 
          & $U^{c}_{1\ 2}\ D^{c}_{1,2}\ E^{c}_{1,2}\ N^{c}_{1,2}$
          & $H^{u}_{1,2}\ H^{d}_{1,2}$
          & $Q_{3}\ L_{3}$
          & $U^{c}_{3}\ D^{c}_{3}\ E^{c}_{3}\ N^{c}_{3}$
          & $H^{u}_{3}\ H^{d}_{3}$\\ \hline
 $Q_{6}$  & ${\bf 2}$
          & ${\bf 2}^{'}$
          & ${\bf 2}^{'}$
          & ${\bf 1}^{'}$
          & ${\bf 1}^{'''}$
          & ${\bf 1}^{'''}$\\ \hline
\end{tabular}
\label{table7}
\end{center}
\end{table}

\subsubsection{Computation of anomaly coefficients}
The $A_{Q_{6}}$ transformation properties of $Q$ and $L$ are given by
\begin{eqnarray}
 \Psi_{\alpha=1-3}\rightarrow 
   \left(\begin{array}{ccc}
      1/2        & \sqrt{3}/2 & 0 \\
     -\sqrt{3}/2 & 1/2        & 0 \\
      0          & 0          & 1
   \end{array}\right)_{\alpha\beta}
 \Psi_{\beta=1-3}.\label{eq,23}
\end{eqnarray}
Because the determinant of this matrix is equal to $1$, $Q$ and $L$ do 
not contribute to the anomaly for this transformation.
On the other hand, $U^{c},D^{c},E^{c},N^{c},H^{u}$ and $H^{d}$
transform as
\begin{eqnarray}
 \Psi_{\alpha=1-3}\rightarrow 
   \left(\begin{array}{ccc}
     -1/2        & \sqrt{3}/2 & 0 \\
     -\sqrt{3}/2 & -1/2       & 0 \\
      0          & 0          & e^{i\frac{2\pi}{6}3}
   \end{array}\right)_{\alpha\beta}
 \Psi_{\beta=1-3}.\label{eq,24}
\end{eqnarray}
These supermultiplets contribute $3$ to the anomaly coefficients, 
and the anomaly coefficients are found to be
\begin{eqnarray}
 &&2{\cal A}_{3}=0\cdot2+3+3=6\ ({\rm mod}\ 6), \\
 &&2{\cal A}_{2}=0\cdot3+0+3+3=6\ ({\rm mod}\ 6).
\end{eqnarray}
As we discussed in \S4, these coefficients do not
contribute to the anomaly.

In the same way, we can compute the anomaly for the transformation
corresponding to $B_{Q}$.
Under this transformation, $Q$ and $L$ transform as
\begin{eqnarray}
 \Psi_{\alpha=1-3}\rightarrow 
   \left(\begin{array}{ccc}
      i &  0 & 0 \\
      0 & -i & 0 \\
      0 &  0 & e^{i\frac{2\pi}{4}2}
   \end{array}\right)_{\alpha\beta}
 \Psi_{\beta=1-3},
\end{eqnarray}
and $U^{c},D^{c},E^{c},N^{c},H^{u}$ and $H^{d}$ transform as
\begin{eqnarray}
 \Psi_{\alpha=1-3}\rightarrow 
   \left(\begin{array}{ccc}
      1 & 0  & 0 \\
      0 & -1 & 0 \\
      0 & 0  & e^{i\frac{2\pi}{4}1}
   \end{array}\right)_{\alpha\beta}
 \Psi_{\beta=1-3}.
\end{eqnarray}
The anomaly coefficients are found to be
\begin{eqnarray}
 &&2A_{3}=2\cdot2-1-1=2\ ({\rm mod}\ 4),\\
 &&2A_{2}=2\cdot3+2-1-1=6\ ({\rm mod}\ 4).
\end{eqnarray}
Therefore the $Z_{4}$ part is anomalous.
\subsubsection{Cancellation of anomalies and gauge coupling unification}
\begin{figure}[p]
 \begin{center}
  \includegraphics[width=13cm,clip]{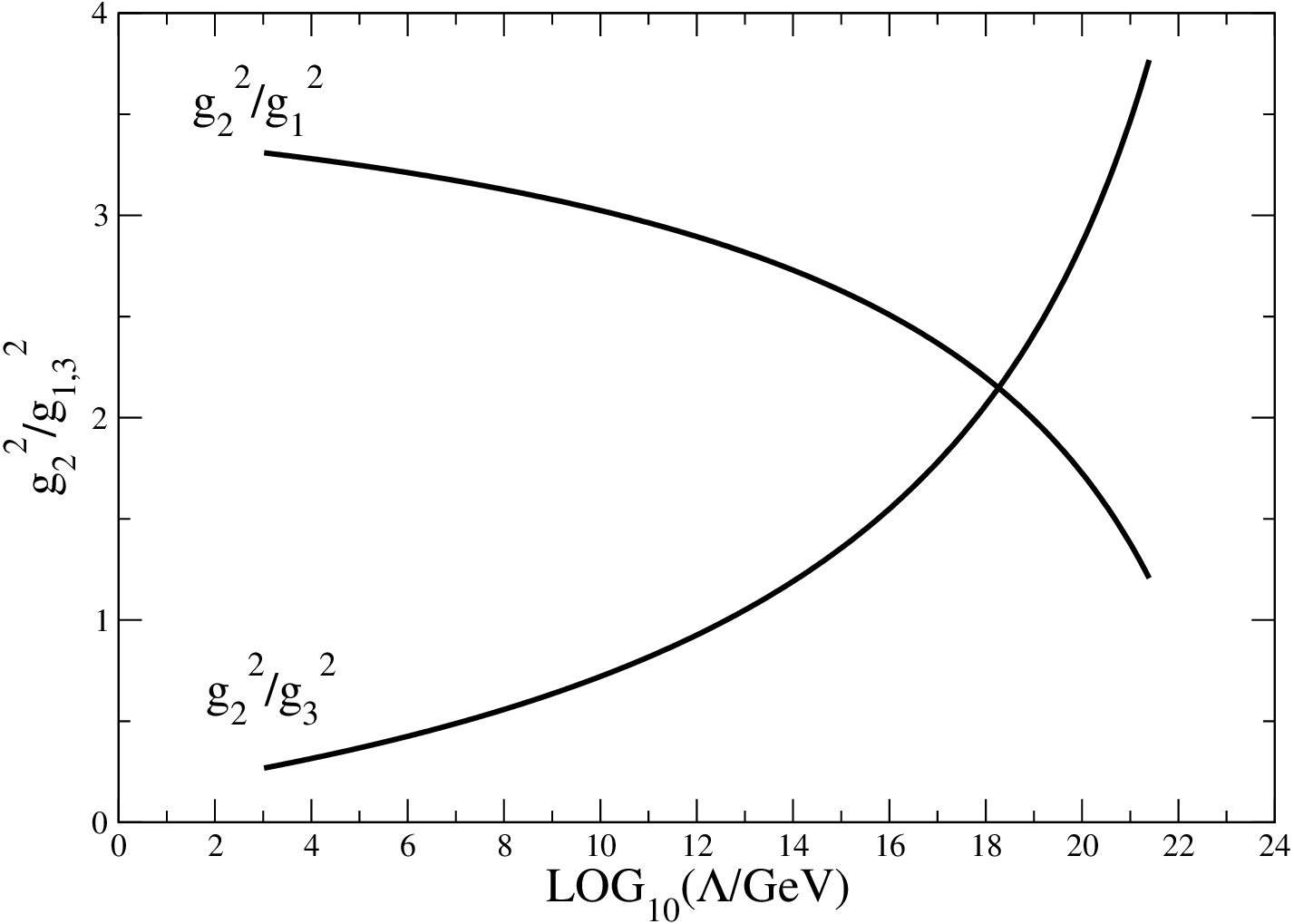}
  \caption{\footnotesize
The ratio $g_2^2/g_1^2$ (upper curve) and $g_2^2/g_3^2$ (lower curve)
as functions of the energy scale.}
  \label{fig,3}
  \vspace{0.5cm}
  \includegraphics[width=13cm,clip]{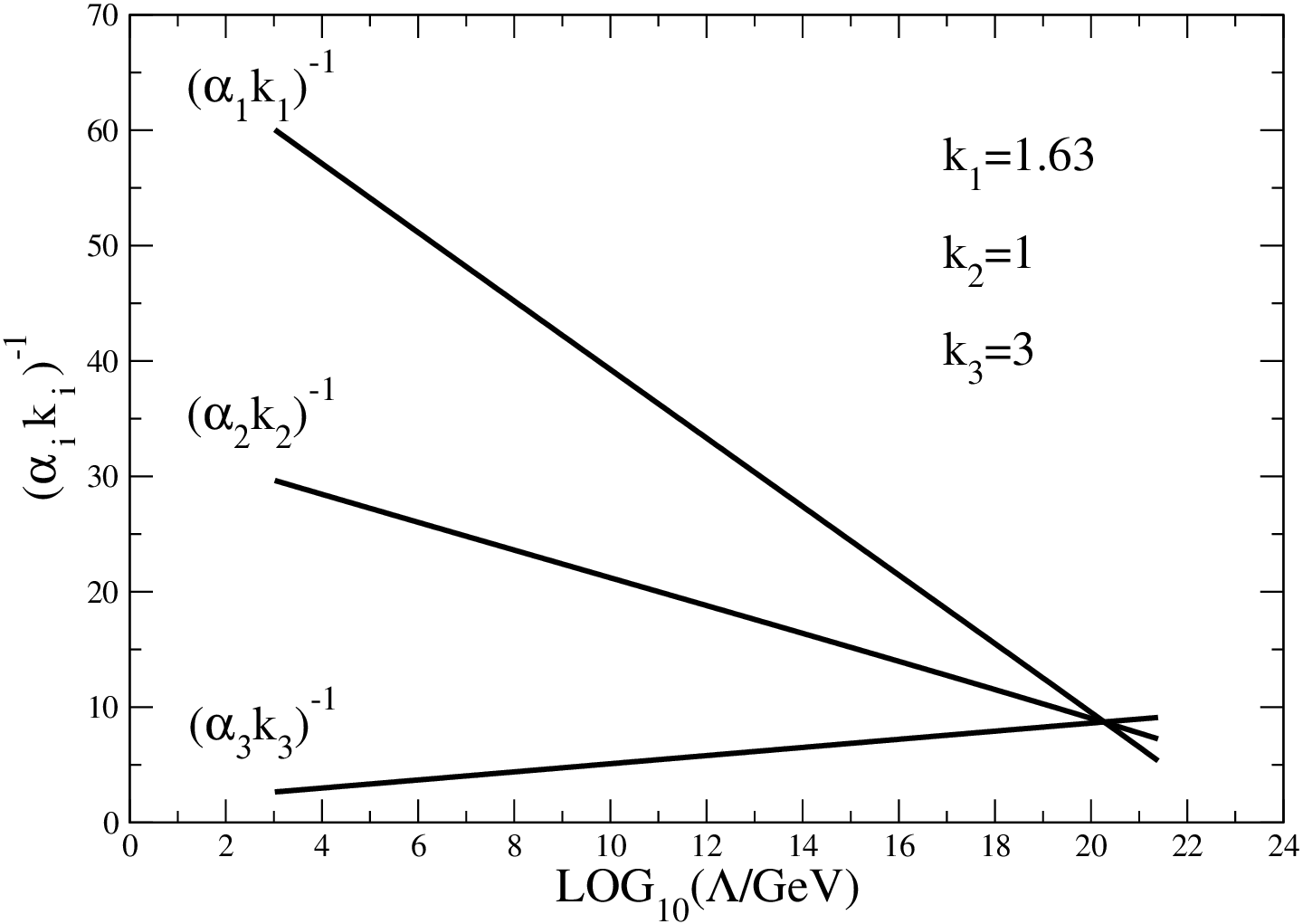}
  \caption{\footnotesize
The running of $(\alpha_1 k_1)^{-1}$, $(\alpha_2 k_2)^{-1}$
and $(\alpha_3 k_3)^{-1}$ with $k_{3}=3, k_{2}=1$ and $ k_{1}\simeq1.63$.
The unification scale is $10^{20}$ GeV.}
  \label{fig,4}
 \end{center}
\end{figure}
As we have seen in the above subsections, the $Z_{4}$ part is anomalous, 
while the $Z_{6}$ part is anomaly-free.
This anomaly is canceled by the GS mechanism if the relation
\begin{eqnarray}
 \frac{1\ ({\rm mod}\ 2)}{k_{3}}=\frac{3\ ({\rm mod}\ 2)}{k_{2}}
\end{eqnarray}
is satisfied.
For example, this is possible if the values of the Kac-Moody levels 
satisfy $k_{3}=k_{2}$ or $k_{2}=1(3)$ and $k_{3}=3(1)$.
Figure \ref{fig,3} displays the ratio of gauge couplings.
For the case $k_{3}=k_{2}$ and the case $k_{2}=3$ and $k_{3}=1$, the
unification point of the $SU(3)_{C}$ and $SU(2)_{L}$ gauge couplings 
is far lower than the Planck scale.
In the case $k_{2}=1$ and $k_{3}=3$, the $SU(3)_{C}$ and $SU(2)_{L}$ 
gauge coupling constants can be unified at a scale slightly higher than 
the Planck scale.
In this case, it is also possible to unify $U(1)_{Y}$ at the same point 
if we assume $k_{1}\simeq 1.63$.
Figure \ref{fig,4} plots the running of the gauge couplings in the case 
$k_{3}=3, k_{2}=1$ and $k_{1}\simeq1.63$.

\section{Summary}
In this paper, we have investigated the anomaly of discrete symmetries 
and their cancellation mechanism.
We have seen that the anomalies of discrete symmetries can be defined
as the anomalous Jacobian of the path-integral measure, and that if
we assume the anomalies are canceled by the GS mechanism, the ordinary 
conditions of gauge coupling unification can be changed.
For the discrete abelian baryon number and lepton number symmetries
in the MSSM with the see-saw mechanism, we find that the ordinary
unification conditions of the gauge couplings are inconsistent with the 
GS cancellation conditions, and that the existence of 
three pairs of  $SU(2)_L$ Higgs  doublets
is a  possible solution to
satisfy the GS cancellation conditions
and  the unification conditions simultaneously.

We have investigated the cases of several recently proposed 
supersymmetric models with a non-abelian discrete family symmetry.
In the examples considered in this paper, the gauge couplings do not
exactly meet at the Planck scale, but we think that the examples
suggest the right direction.
If we take into account the threshold corrections at
$M_{PL}$, for instance, the conditions could be exactly satisfied.

\end{document}